\documentclass[prb,preprint]{revtex4-1}

\usepackage{float}
\usepackage{amsmath,amsfonts,amsthm} 
\usepackage{dcolumn}
\usepackage{graphicx}
\usepackage{amssymb}
\usepackage{multirow}
\usepackage{setspace}
\usepackage{needspace}

\newcommand{\degree}{\ensuremath{^\circ}}

\draft 

\begin{document}

\title{Inexpensive LED-based spectrophotometer for analyzing optical coatings} 

\author{Kayla Hardie}
\email[]{khardie@uwaterloo.ca}
\author{Sascha Agne}
\author{Katanya B. Kuntz}
\author{Thomas Jennewein}
\affiliation{Department of Physics and Astronomy and Institute for Quantum Computing, University of Waterloo, Waterloo, Ontario N2L 3G1, Canada}

\date{\today}

\begin{abstract}

Optical coatings are widespread in everyday life, from camera lenses to glasses, to complex optics experiments. A simple, reliable device that can quickly and inexpensively analyze optical coatings is a valuable laboratory tool. Such a device can identify unknown or mislabelled optics, and characterize the transmission spectra of optical elements used in an experiment. We present the design and characterization of a LED-based spectrophotometer, and demonstrate its ability to identify different optical coatings. Our approach uses ten LEDs that cover a spectrum from 365\,nm to 1000\,nm. A small servomotor and microcontroller rotates a LED board to sequentially position each LED over an optical sample, and the transmitted light corresponding to each LED is measured with a silicon photodetector. The device is automated, portable, inexpensive, user-friendly and simple to build.

\end{abstract}


\maketitle 

\section{Introduction}

An optical coating is composed of one or several thin film layers, each with carefully chosen thicknesses and indexes of refraction. The coatings are deposited onto the surface of an optical element, such as a lens or wave plate, to alter the reflection and transmission for particular wavelengths.  Coatings are widespread in everyday life, from camera lenses to banknote security, and are equally important in optics manufacturing and research. For instance, anti-reflective coatings are used to minimize the four percent transmission loss at air-glass interfaces. \cite{Hecht2001} In a laboratory setting, the ability to analyze the performance of an optical coating for a given spectrum is extremely useful. Such a device helps to characterize a custom coating, identify mislabelled optics, or assess whether a coating has degraded over time. 

Traditional spectrophotometers for optical coating testing consist of a white light source and a monochromator to select a narrow band of wavelengths. A typical commercial device, such as the F10-AR thin film analyzer from Filmetrics, can precisely measure the spectra of an anti-reflection coating, its thickness and material composition. Unfortunately, these devices are quite expensive, ranging from tens to hundreds of thousands of US dollars. 

Recently, there has been interest in replacing costly, complicated spectrophotometers with simpler devices that use light emitting diodes (LEDs) as the selective wavelength source. \cite{Dasgupta1993,Dasgupta2003,Cantrell2003,Yeh2006,Tavener2007,OToole2008,Macka2014,Bui2015} LEDs have several advantages, most notably their small size, negligible warm up time, low cost and low power requirements. They have numerous applications, such as in absorbance measurements of chemical components \cite{Flaschka1973,Klughammer1990,Hauser1995,Fonseca2004} or in fluorometry.  \cite{Szmacinski2000,Herman2001} One of the first reported analytical LED devices used a red LED and a phototransistor to make a photometric instrument for determining the chemical concentration of an unknown solution via absorbance measurements. \cite{Flaschka1973}

Since LEDs have a fixed wavelength, changing the wavelength usually requires physically replacing the LED with another. To overcome the limitation of a fixed emission band, the use of a switchable multi-channel LED array was introduced. \cite{Klughammer1990,Hauser1995,Schnable1998,Fonseca2004,Masi2011,Lee2012} There are various ways to direct light from a multi-channel array towards the test sample. Hauser et al. \cite{Hauser1995} coupled the light from seven LEDs into a seven channel optical fibre coupler, and used an electronic switch to individually activate each LED. A wider spectrum of wavelengths made it possible to identify several chemical elements, instead of only one or two, in an unknown solution using standard colorimetric methods. Lee et al. \cite{Lee2012} demonstrated another way to make a switchable LED array by using a DC servomotor to rotate the appropriate LEDs towards the object under forensic investigation. However, all of these devices were used in absorbance measurements to identify concentrations of specific chemicals, and have not been applied to analyze optical coatings. 

We have built a simple, inexpensive optical coating analyzer that measures the transmission spectrum of an optical element, such as a lens, to analyze its coating as a function of wavelength. Our approach uses ten LEDs and a silicon photodetector to measure the transmission spectrum of a passive optic for a discrete set of wavelengths. The device costs under three hundred US dollars to build, and only takes a few days to assemble. The design is similar to Lee et al. \cite{Lee2012} in that we use a small servomotor to rotate a LED board to sequentially position each LED over a test optic, and measure the transmitted light as a function of LED wavelength. We also introduce a simple calibration method to compensate for the extended emission spectra of some LEDs. This correction technique does not require {\it{a priori}} knowledge about the optical element being studied. To our knowledge, this is the first application of an LED-based spectrophotometer to analyze and identify optical coatings.

\section{The Device}

\begin{figure}
\centering{\includegraphics[height=6.7 cm]{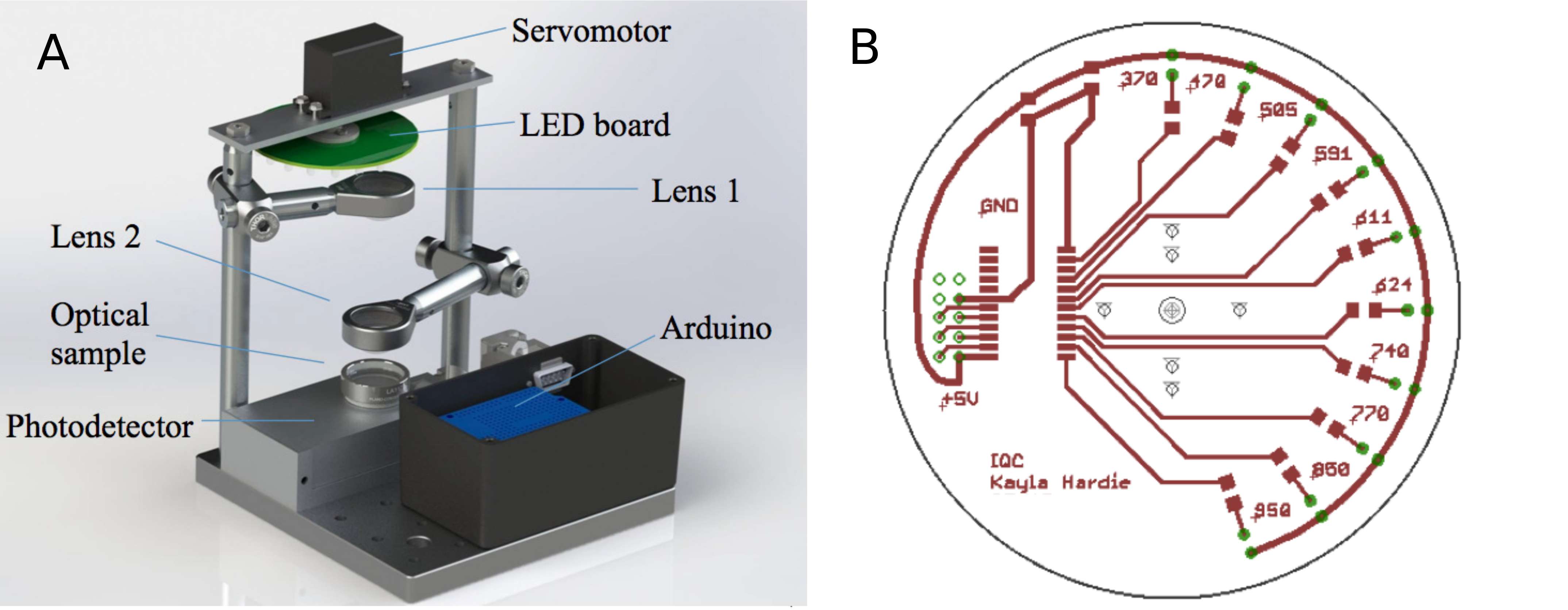}}
\caption{A) {\it{SolidWorks}} \cite{SolidWorks} render of our LED-based spectrophotometer. The LEDs and photodetector are used to capture the transmission spectrum of an optical sample placed above the photodetector. A small servomotor was used to rotate the LED board, which sequentially positioned each LED above the alignment optics (lenses), optical sample and photodetector.  A microcontroller (an Arduino) was used to control and power the entire device, and captured the photodetector signal corresponding to transmitted light from each LED. B) LED board circuit layout generated using {\it{Eagle}} \cite{Eagle} design software.}
 \label{fig:Device}
\end{figure}

Our device in Fig. \ref{fig:Device} uses ten LEDs and a silicon photodetector to measure the transmission spectrum of a passive optical element, such as a lens or optical filter, for a discrete set of wavelengths defined by the emission spectrum of each LED. Ten LEDs were chosen, ranging from ultraviolet (370\,nm) to near-infrared (950\,nm), to give a wide range of wavelengths that covers the most commonly used coating types in optics laboratories. Most multi-channel LED devices couple all the light into a single optical filter, and use an electronic switch to activate each LED individually. We chose to mount the LEDs on a rotating circular printed circuit board for simplicity, which was driven by a servomotor. A servomotor was chosen because of its positional awareness, and since additional motor drivers were not required. A 4-to-16 line decoder/demultiplexer was then used to sequentially activate each LED once positioned above the test optic and photodetector. Two lenses were placed between the LED and test optic, as shown in Fig. \ref{fig:Device}A, to collect and collimate the LED light through the optic and onto the photodetector. An enclosure with an aperture was placed over the photodetector to block stray light. Employing collimation optics was important as it helps to counteract the dispersing effect caused by the LED dome package. Since the final transmission spectrum of the test optic was found by taking a ratio relative to a reference measurement, the optical coatings of these intermediate optics do not affect the measured transmittance of the sample. As a result, calibration of the intermediate optics and photodetector response was not necessary. A light-tight box was placed over the entire device during measurement to block ambient light.

The entire system was controlled and powered by a ATmega328 microcontroller (Arduino Uno). The Arduino was chosen because of its low cost, prototyping ease and well-documented use. It can be powered by a computer through a USB connection, which also served as a line of communication between the user and the device for operation and data collection. We developed a user interface that controls the device to perform calibration measurements, and to characterize the optical coating of a sample (see Supplementary Materials for software code \cite{Github}).

\subsection{The physical arrangement}
We chose a vertical arrangement for the source and optics (see Fig. \ref{fig:Device}) so that the sample can be placed directly on top of the photodetector enclosure, which limits the distance between the test optic and detector. This arrangement reduced the possibility for beam aberrations, or the detector becoming misaligned due to beam steering by the optical sample. A simple foam-covered platform was designed to protect the surface of an unmounted sample. The platform had an aperture to allow light to pass straight through onto the photodetector.

Additionally, both the photodetector and microcontroller enclosures were mounted on a breadboard in adjustable slots. Thus, the position of either enclosure can be easily adjusted if necessary, which allowed for overall flexibility in future designs. The two collimation lenses also provided more degrees of freedom for alignment purposes. All of the electronics were connected using D-sub serial connectors, and the motor was connected using an M3 coaxial connector. These connectors made the electronic connections more robust and accessible. Finally, we placed a light-tight box over the entire system to block ambient light during measurements. This box ensured that the dark noise level of the silicon photodetector was low enough to not affect the measurement accuracy. 

\subsection{The LED board}

Light-emitting diodes are non-linear devices that exhibit the typical logarithmic current versus voltage characteristics of a diode. The light originates from the semiconducting material contained in a reflective cup (or dome), which is connected to one of the connecting leads. The second lead is also connected to the semiconducting material from the top by a very thin wire. 

\begin{table}[]
\centering
\begin{tabular}{|c|c|c|c|c|c|c|c|}
\hline
LED name & $\lambda_{\mathrm{meas}}$ (nm) & $\Delta \lambda_{\mathrm{meas}}$ (nm) & $R_i$ ($\Omega$) & PWM & $\theta$ ($\degree$) & $\sigma_{\mathrm{ref}}$ (\%) & $\sigma_{\mathrm{lens}}$ (\%) \\ \hline
370\,nm & 378.7        & 11.0                & 120   & 0   & 2         & 0.03   & 0.02      \\
 & $600^*$ & $40^*$ & $-$ & $-$ & $-$ & $-$ & $-$ \\ \hline
470\,nm & 477.6        & 37.4                & 91    & 252 & 20        & 0.1    & 0.08      \\ \hline
505\,nm & 510.4        & 37.8                & 91    & 250 & 40        & 0.2    & 0.2       \\ \hline
591\,nm & 592.4        & 7.0                 & 147   & 250 & 57        & 0.2    & 0.1       \\ \hline
611\,nm & 612.9        & 29.6                & 158   & 247 & 76        & 0.2    & 0.1       \\
      & 864.9        & 66.8                & $-$   & $-$ &  $-$     &  $-$    & $-$         \\ \hline
624\,nm & 629.0        & 14.4                & 147   & 253 & 94        & 0.09   & 0.3       \\ \hline
740\,nm & 737.0        & 26.9                & 160   & 254 & 114       & 0.2     & 0.3         \\ \hline
770\,nm & 764.7        & 25.0                & 174   & 254 & 136       & 0.1   & 0.08      \\ \hline
850\,nm & 850.3        & 18.5                & 357   & 253 & 154       & 0.2    & 0.09      \\ \hline
950\,nm & 937.3        & 37.6                & 187   & 251 & 174       & 0.1     & 0.09      \\ \hline
\end{tabular}

\caption{Summary of LEDs used in our device. The LED names are the centre wavelengths specified by the manufacturer. The measured centre wavelength ($\lambda_{\mathrm{meas}}$) and full width at half maximum ($\Delta \lambda_{\mathrm{meas}}$) were found by curve fitting the spectrometer data shown in Fig. \ref{fig:LEDspectrum}A. $R_i$ refers to the LED-current limiting resistors, where $i = 1\dots10$ corresponds to the labelling convention used in the final circuit diagram of the device (see Appendix, Fig. \ref{fig:WiringDiag}). The pulse width modulation (PWM) and servomotor angle ($\theta$) settings correspond to the final device configuration. $\sigma_{\mathrm{ref}}$ and $\sigma_{\mathrm{lens}}$ are the relative standard deviation (ratio of standard deviation to mean value) of ten trials per LED taken of both the reference measurements and measurements with a sample lens, respectively. Note that two of the LEDs (370\,nm and 611\,nm) have two measurable emission peaks. \:$^*$ Assumed $\lambda_{\mathrm{meas}}$ and  $\Delta \lambda_{\mathrm{meas}}$ based on available data in Fig. \ref{fig:LEDspectrum}A.}
\label{tab:LEDs}
\end{table}

Specific LEDs for our device were chosen based on their luminosity, wavelength, packaging, bandwidth, and cost. The relevant attributes for each LED are summarized in Table \ref{tab:LEDs}. The wavelengths were chosen to fall within the detection bandwidth of a silicon photodiode (350\,nm$-$1100\,nm). All of the LEDs were less than one US dollar, with the exception of the UV LED, which was around US\$15. Ideally, we wanted LEDs that were inexpensive and bright enough to be detected by our photodetector. Since the photodetector measures a sum of the entire emission spectrum of each LED, we tried to find LEDs with reasonably narrow bandwidths to achieve coating measurements that correspond to a small wavelength range. We chose 5mm, radial, through-hole LEDs as it was easier to prototype with this type of LED package.  

A constant current source is necessary to operate an LED in a linear mode. This can be approximated by placing a current limiting resistor in series with the LED. The resistance value can be calculated using Ohm's Law by taking the ratio between the voltage rail (+5\,V) and desired LED forward current, which should be less than the specified maximum value for each LED.

The LEDs were controlled by a 4-to-16-line decoder/demultiplexer (74HC154) and a microcontroller. The 74HC154 was chosen because it can selectively turn on a single LED. According to the decoder's data sheet, every unique combination of four bits will send one output to LOW, while the rest of the outputs remain on HIGH. Since the LEDs were connected to the +5\,V rail of the microcontroller and decoder, when one of the outputs was on LOW, current only flows to a particular LED, which caused it to turn on. The decoder was a pin efficient choice as it only required six microcontroller pins (four address inputs and two enable inputs) instead of 16 signals to control up to 16 LEDs. 

\begin{figure}
\centering{\includegraphics[height=4.6 cm]{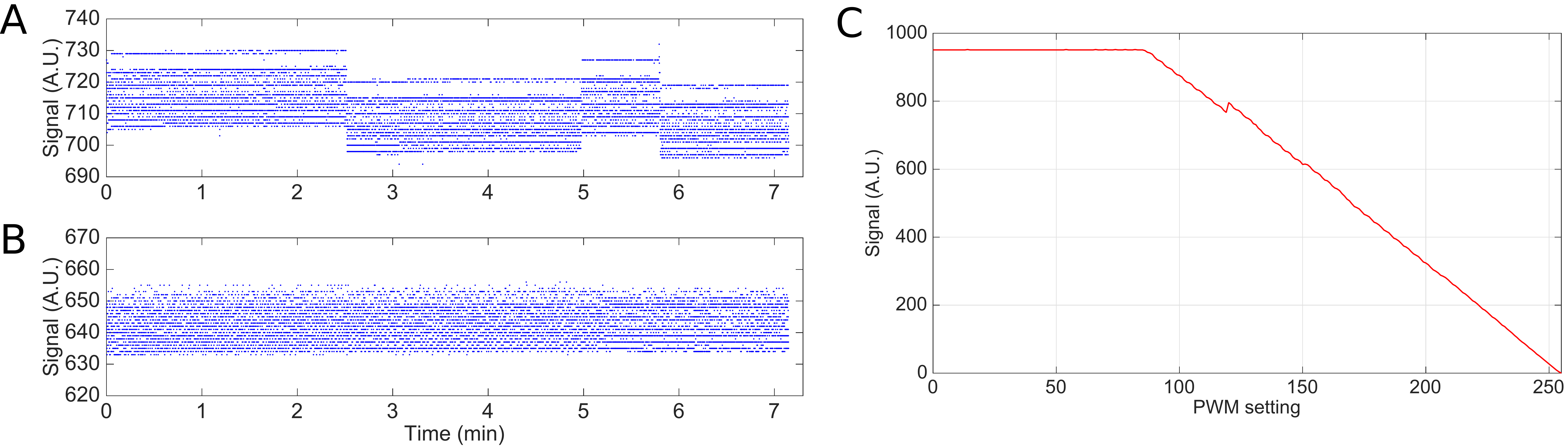}}
\caption{A) Photodetector signal from one LED illustrating the instability over time caused by leaving the servomotor on active (ie. before the {\it{detach()}} command was implemented). B) Photodetector signal from same LED after implementation of the {\it{detach()}} command to effectively deactivate the servomotor, which eliminated the continuous micro-adjustments to servomotor position. C) Example of how photodetector signal level from one LED can be adjusted via PWM setting.}
 \label{fig:MotorPWM}
\end{figure}

The LED board was attached to a small servomotor, which rotated each LED into position over the collimating optics and photodetector. We used a 3000 - Hitec HS-422 Deluxe Servo, which only required 4.8\,V$-$6\,V to power, and can therefore be powered by the microcontroller. The microcontroller also controlled the position of the servomotor using pulse width modulation (PWM). The motor continuously monitored the pulse width signal from the microcontroller to determine its position. As a result, if the motor was left connected to the pulse width signal during measurements (i.e. left on active), it would continue micro-adjusting its position. These adjustments created a varying photodetector signal even while the motor was supposed to be stationary, as shown in Fig. \ref{fig:MotorPWM}A. We corrected this instability in the microcontroller code by using the {\it{detach()}} command in the Arduino servomotor library, which effectively disassociated the motor with the pulse width pin. This had the resulting effect of stabilizing the photodetector signal while the motor was stationary, as shown in Fig. \ref{fig:MotorPWM}B.

The 4-to-16-line decoder also provides an efficient way to adjust the brightness of each LED. The decoder required that its two enable inputs were set to LOW to operate. Setting the two enable inputs to LOW was essentially activating two separate `on' switches. We can fine-tune the effective brightness of each LED by setting one of the enable pins to LOW, and using PWM to manipulate the other pin. An example of the photodetector's response as the PWM setting was tuned for one LED is shown in Fig. \ref{fig:MotorPWM}C. Coarse adjustment of LED brightness was accomplished by selecting appropriate resistors for each LED. The LEDs should be bright enough to register within the photodetector's dynamic range with enough clearance to stay well above dark noise but not saturate the detector. It can be seen in Fig. \ref{fig:MotorPWM}C that the photodetector was saturated for a PWM setting below 90. The small signal discontinuity visible was caused by motor instability before the {\it{detach()}} command was implemented.The final PWM settings (listed in Table \ref{tab:LEDs}) were selected once the collimation setup was finalized, and resulted in signal values that ranged between 45\%$-$60\% of the maximum reading.

\subsection{The collimation optics}

Two plano-convex lenses were used, both with a focal length of 25.4\,mm, to collect and collimate the LED light towards the optical sample and photodetector. The focal length was chosen based on the constraints of collimating the light within a short distance (approximately 10\,cm), and passing through the small aperture of the photodetector enclosure. The height of the top lens (Lens 1 in Fig. \ref{fig:Device}A) was positioned as close as possible to the LEDs (4\,mm), while still clearing the rotating components on the LED board. The role of the upper lens was to collect as much light from the LED as possible before it disperses, and direct it down towards the photodetector. The second lens (Lens 2 in Fig. \ref{fig:Device}A) is approximately 1.3\,cm above the optical sample, and was adjusted to focus  the light onto the test optic and into the photodetector enclosure. It was important to ensure the light was not clipping on the enclosure's aperture, as that would cause beam aberrations and possibly affect the results.

\subsection{The photodetector}
The LED light was measured using a silicon photodiode. The photodetector circuit (see Appendix, Fig. \ref{fig:WiringDiag}) used a transimpedance operational amplifier to convert the photocurrent from the photodiode into an amplified voltage. A LTC1050 operational amplifier was chosen due to its low noise and input bias current, and it can be powered by a single +5\,V power supply, making it easy to power using the microcontroller. The photodetector signal was connected to the microcontroller's analog input which digitally sampled the signal at a rate of 20\,Hz. The photodetector signal was averaged over 50 samples and stored in the software. The amplifier circuit had a slow time constant of approximately 0.25\,s in order to average over the PWM of the LEDs. The photodetector circuit was placed inside an enclosure to shield the circuit from high-frequency environment noise.

\section{Device Operation}
The general procedure for characterizing an optical sample was to first capture a reference transmission spectrum by measuring the photodetector signal corresponding to each LED without a test optic in place. This procedure was then repeated with a test optic. A ratio was taken between the two transmission spectra, which calculated the transmittance of the test optic as a function of LED wavelength. 

In order to identify the optimal servomotor angle position for each LED, a series of detector readings were captured for a range of angles between $0\degree-180\degree$ without a test optic present. A calibration program stepped through angles around the estimated optimal angle for each LED, and captured the photodetector averaged signal for each angle. The results from these measurements are shown in the Appendix, Fig. \ref{fig:MotorAngle}. The servomotor stepping in increments of $1\degree$ provided a fine enough angular resolution to locate a well-defined maximum in the photodetector signal for each LED. 

\subsection{Spectral characterization and calibration}

\begin{figure}
\centering{\includegraphics[height=13 cm]{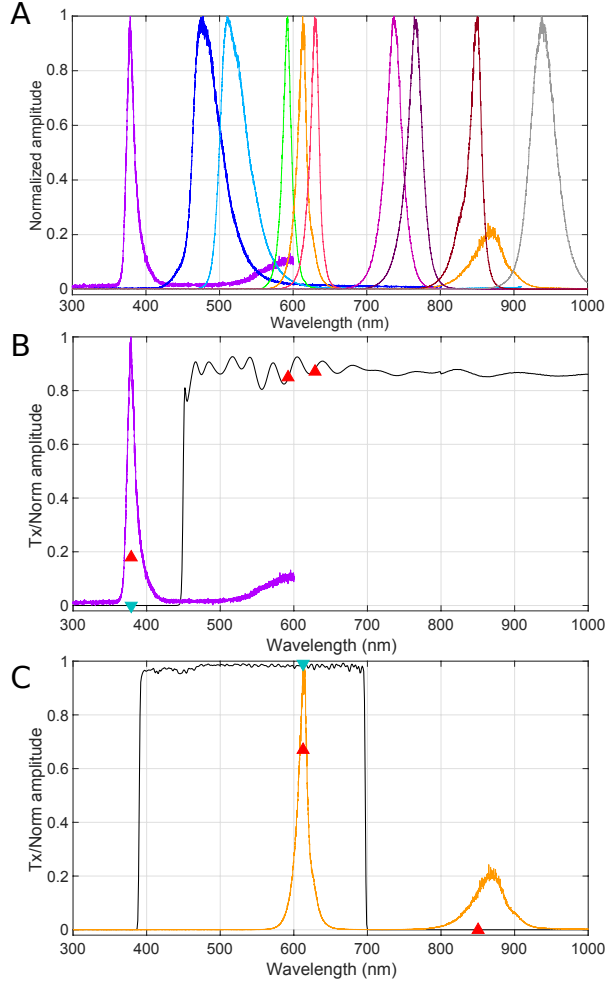}}
\caption{A) Measured emission spectra of our ten LEDs. From right to left: 370\,nm, 470\,nm, 505\,nm, 591\,nm, 611\,nm, 624\,nm, 740\,nm, 770\,nm, 850\,nm, and 950\,nm LEDs. Note there are two measurable emission peaks in the 370\,nm and 611\,nm LED spectra. B) Filter data used for correcting double peaked emission spectrum of 370\,nm LED (purple solid line). Thorlabs \cite{Thorlabs} transmission data for FEL0450 longpass filter is shown (black solid line), as well as uncorrected transmittance data for 370\,nm, 591\,nm, and 624\,nm LEDs (red upward-pointing triangle), and corrected 370\,nm LED data (green downward-pointing triangle). C) Filter data used for correcting double peaked emission spectrum of 611\,nm LED (orange solid line). Thorlabs \cite{Thorlabs} transmission data for FESH0700 shortpass filter is shown (black solid line), as well as uncorrected transmittance data for 611\,nm and 850\,nm LEDs (red upward-pointing triangle), and corrected 611\,nm LED data (green downward-pointing triangle).}
 \label{fig:LEDspectrum}
\end{figure}

We measured the emission spectra of the ten LEDs as part of calibrating our device before use. Figure \ref{fig:LEDspectrum}A shows the measured LED emission spectra performed with a commercial liquid nitrogen spectrometer ({\it{SpectraPro 2750}}). Light from each LED was coupled into a multi-mode fibre by replacing the top collimation lens with a fibre coupler. The fibre was then connected to the spectrometer, and the LED brightness was adjusted using the PWM setting to avoid saturating the spectrometer. Note there are clearly two emission peaks for the 370\,nm and 611\,nm LEDs, which is probably related to their chip construction. \cite{Neamen2002} It is possible that the other LEDs also have secondary emission peaks but our spectrometer was not sensitive enough to detect them. 

The double emission peak behaviour of certain LEDs is a well-known phenomenon, \cite{Neamen2002} and is usually mitigated via additional optical filters to eliminate the unwanted spectrum. \cite{Hauser1995} However, adding additional filters would increase the cost and complexity of our device. Therefore, we developed a simple calibration model to correct for the known double emission peaks from the 370\,nm and 611\,nm LEDs. This correction technique does not require {\it{a priori}} knowledge about the optical element being studied. 

We used the filter data shown in Fig. \ref{fig:LEDspectrum}B and \ref{fig:LEDspectrum}C to estimate the approximate ratio of signal that corresponds to the main and secondary peaks. The captured photodetector signal is a measure of photocurrent, which is proportional to the number of photons incident on the photodiode. Therefore, we can express the total measured photocurrent as a sum of contributions from the main and secondary peaks as
\begin{eqnarray}
\label{eqn:Iref}
I_{\mathrm{ref}} &=& I_1 + I_2 ,	 \\ \label{eqn:Iopt}
I_{\mathrm{opt}} &=& I_1T_1 + I_2T_2 ,   
\end{eqnarray}
where $I_{\mathrm{ref}}$ is the total measured reference photocurrent for one LED, $I_{\mathrm{opt}}$ is the total measured photocurrent for the sample optic for one LED, and $I_1$ and $I_2$ are the photocurrent contributions from the main and secondary emission peaks, respectively. $T_1$ and $T_2$ are the transmittances for the wavelength range around the main and secondary emission peaks, respectively. The spectrometer data shows that the main peak for the 370\,nm LED is centred at 378.7\,nm, and we assumed the secondary emission peak extends between 520\,nm$-$700\,nm and centred at 600\,nm. While the main and secondary peaks for the 611\,nm LED are centred at 612.9\,nm and 864.9\,nm, respectively. Therefore, we can estimate $I_1$ and $I_2$ using the FEL0450 data (Fig. \ref{fig:LEDspectrum}B) for the 370\,nm LED , and the FESH0700 data (Fig. \ref{fig:LEDspectrum}C) for the 611\,nm LED. We used the average measured transmittance from the 591\,nm and 624\,nm LEDs to calculate $T_2$ for the FEL0450 filter. Similarly, we used the transmittance data from the 850\,nm LED to estimate $T_2$ for the FESH0700 filter. This resulted in $I_1\,=\,0.792I_{\mathrm{ref}}$ and $I_2\,=\,0.208I_{\mathrm{ref}}$ for the 370\,nm LED, and $I_1\,=\,0.677I_{\mathrm{ref}}$ and $I_2\,=\,0.323I_{\mathrm{ref}}$ for the 611\,nm LED. Knowing these quantities allowed us to calculate the corrected transmittance $\mathcal{T}_{1}$ for any optical sample as
\begin{equation} \label{eqn:CorrTx}
\mathcal{T}_1 = \frac{I_{\mathrm{opt}} - I_2T_2}{I_1} .
\end{equation}

\subsection{Error and uncertainty}
To investigate the precision and repeatability of our device, we explored how reference measurements and measurements with a sample optic vary over several trials. Table \ref{tab:LEDs} summarizes the relative standard deviation (ratio of standard deviation to mean value) of ten trials per LED for both the reference and sample optic measurements. This data explores the uncertainty induced by servomotor angle, accuracy of the servomotor, variation in LED brightness due to PWM setting, and optical alignment of the collimating lenses. The largest relative standard deviation for the reference measurements was only $0.2\%$, whilst measurement error with a sample optic only slightly increased to $0.3\%$. Such a small variation between measurements clearly demonstrates the precision of our device.

\subsection{Characterizing optics}

\begin{figure}
\centering{\includegraphics[height=14 cm]{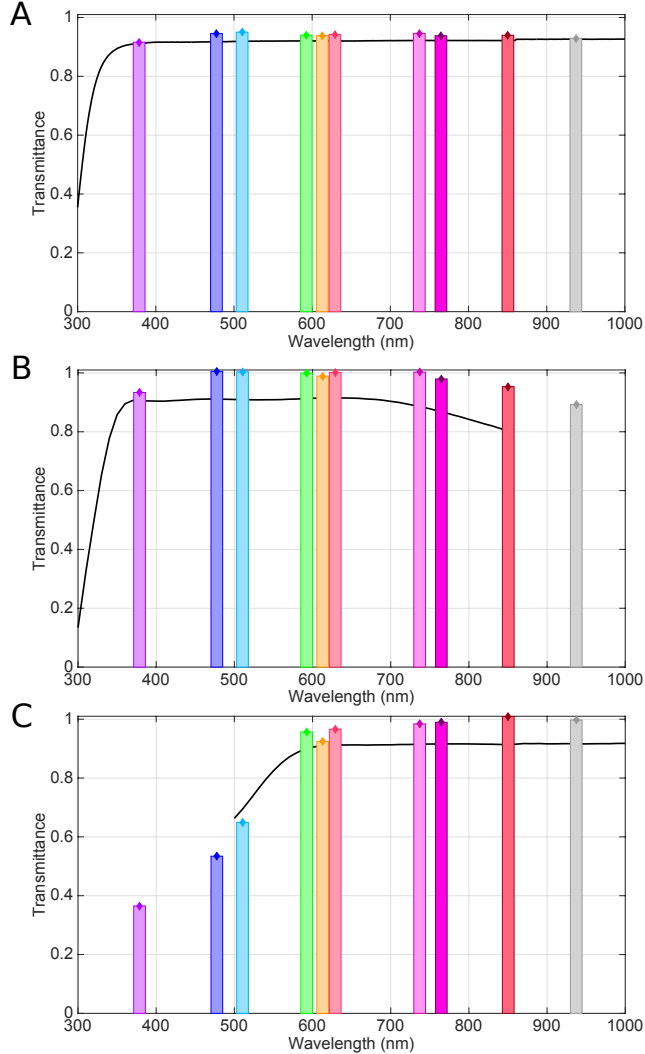}}
\caption{Transmittance spectra for three different anti-reflection coatings corrected for double emission peaks from 370\,nm and 611\,nm LED spectra. The corresponding Thorlabs \cite{Thorlabs} transmission data is shown for comparison (black solid line). A) Uncoated plano-convex lens ($f\,=\,125$\,mm). B) A-coated plano-convex lens ($f\,=\,400$\,mm). C) B-coated plano-convex lens ($f\,=\,150$\,mm). Measurement error is represented by the size of the data points, and the LED bandwidth is represented by the bar width.}
 \label{fig:Lenses}
\end{figure}

\begin{figure}
\centering{\includegraphics[height=13 cm]{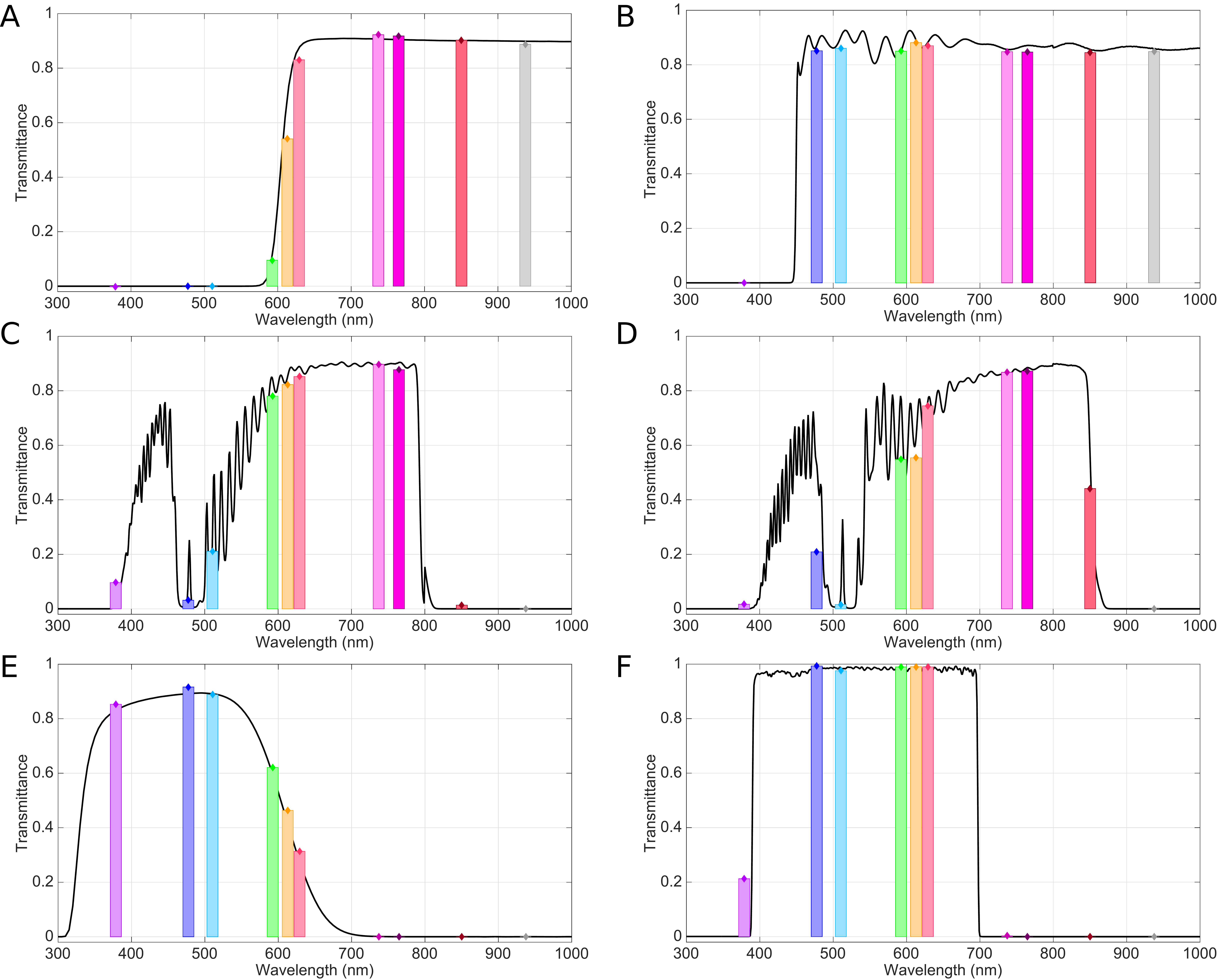}}
\caption{Transmittance spectra for several different optical filters corrected for double emission peaks from 370\,nm and 611\,nm LED spectra. The corresponding Thorlabs \cite{Thorlabs} transmission data is shown for comparison (black solid line). A) FGL610 colored glass longpass filter, B) FEL0450 longpass filter, C) FES0800 shortpass filter, D) FES0850 shortpass filter, E) FGB37 colored glass bandpass filter, F) FESH0700 shortpass filter. Measurement error is represented by the size of the data points, and the LED bandwidth is represented by the bar width.}
 \label{fig:Filters}
\end{figure}

We characterized various optics to demonstrate the ability of our device to distinguish between different types of optical coatings, and to characterize the transmission spectra of several optical filters. Figure \ref{fig:Lenses} shows transmittance data for lenses with three different types of optical coatings: Thorlabs A-coated (optimized between 350\,nm$-$700\,nm), B-coated (optimized between 650\,nm$-$1050\,nm), and uncoated. The 370\,nm and 611\,nm LED data have been corrected using the method previously outlined. Note how the general trend of the measured transmission spectra follows the data provided by Thorlabs. \cite{Thorlabs} Since each LED has a finite bandwidth, as shown in Fig. \ref{fig:LEDspectrum}A, we are not measuring the response from a discrete wavelength, but rather a sum of the entire emission spectrum of each LED. Therefore, the LED bandwidths are represented by the widths of the bars plotted in Fig. \ref{fig:Lenses} and \ref{fig:Filters}.

Figure \ref{fig:Filters} shows the transmittance curves of several optical filters. Our measurements show reasonable agreement with the transmission data provided by Thorlabs. \cite{Thorlabs} In particular, our device is able to measure the noticeable difference in transmittance for the 850\,nm LED between two shortpass filters with slightly different cut-off wavelengths of 800\,nm (Fig. \ref{fig:Filters}C) and 850\,nm (Fig. \ref{fig:Filters}D). Examining these transmittance curves illustrates that our spectrophotometer device can help identify the coating type or transmission spectrum of a passive optic, which is a very useful laboratory tool for identifying unknown or mislabelled optics.

We also investigated how the position of the test optic placed over the photodetector affects transmittance. We measured a transmittance greater than 1 for a few of the LEDs with off-centred lens positions. It is possible that the lens was directing more of the LED light onto the photodetector compared to the amount of light captured during the reference measurements. Lensing effects from the sample optic may also explain the higher than expected transmittance seen in Fig. \ref{fig:Lenses}.

\section{Discussion}

Future improvements to our device include selecting LEDs with a smaller viewing angle and a single emission peak, or using filters to limit the emission spectra. Another improvement is to make directional alignment more consistent, which can be achieved by filing the bottom of each LED package, or by inserting them into mounting holders to sit flush against the LED board. This caused the light to be emitted at a slight angle relative to vertical, which affected alignment and possibly resulted in some light not being captured by the photodetector. Choosing LEDs with a smaller viewing angle would make the measurements less susceptible to slight changes in alignment due to variations in motor angle or position of optical sample between trials.

An additional improvement could be to increase the resistor values for most of the LEDs. We calculated the resistor values based on the specified maximum forward current of each LED. As a result, a majority of the LEDs were too bright, and saturated the photodetector. Therefore, the PWM had to be significantly adjusted to reduce the brightness. A better approach is to apply coarse adjustments to LED brightness via resistor values, and then fine adjustments using the PWM setting. 

Another improvement would be to add more LEDs to obtain more coverage or extend the spectrum. The microcontroller model we used can support up to fourteen LEDs, and there are commercially available LEDs that cover the gaps in our current detection spectrum. Therefore, additional LEDs can be added if we alter the LED board, and replace the servomotor with one that has a full rotational range of $360\degree$. Additional LEDs will improve the coverage of the wavelength range, which should increase the accuracy of our device. It is important to note that LEDs age differently, and the effects of ageing need to be considered in the long term. Therefore, calibration of the device would need to be repeated periodically.

\section{Conclusion}

We have designed and implemented a simple, inexpensive LED spectrophotometer that can successfully distinguish between different types of optical filters and coatings. The relative standard deviation over ten trials remaining well-below $0.5\%$ demonstrates the precision of our device. This laboratory tool is especially useful in an optics lab, as it can quickly identify unknown or mislabelled optics, or estimate the transmission spectrum of an optical element. Obtaining an up-to-date transmission spectrum for optics used in experiments can be crucial, especially if the coating was damaged or degraded over time. To our knowledge, this is the first demonstration of characterizing optical coatings using a simple, automated LED-based spectrophotometer.

\appendix*

\section{Additional material}

\begin{figure}
\centering{\includegraphics[angle=90,origin=c,height=19 cm]{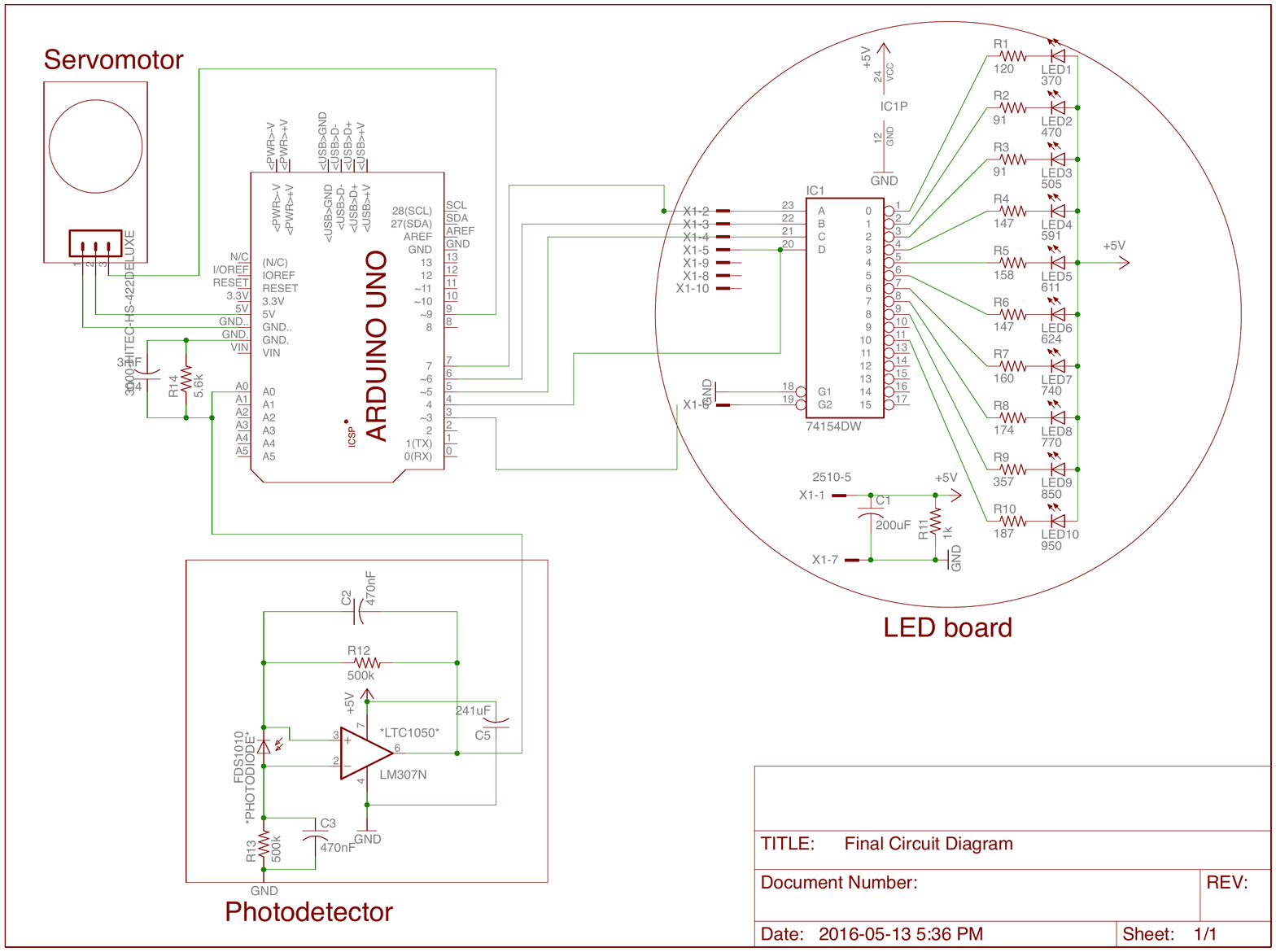}}
\caption{Complete electronic schematic for our LED-based spectrophotometer.}
 \label{fig:WiringDiag}
\end{figure}

\begin{figure}
\centering{\includegraphics[height=7 cm]{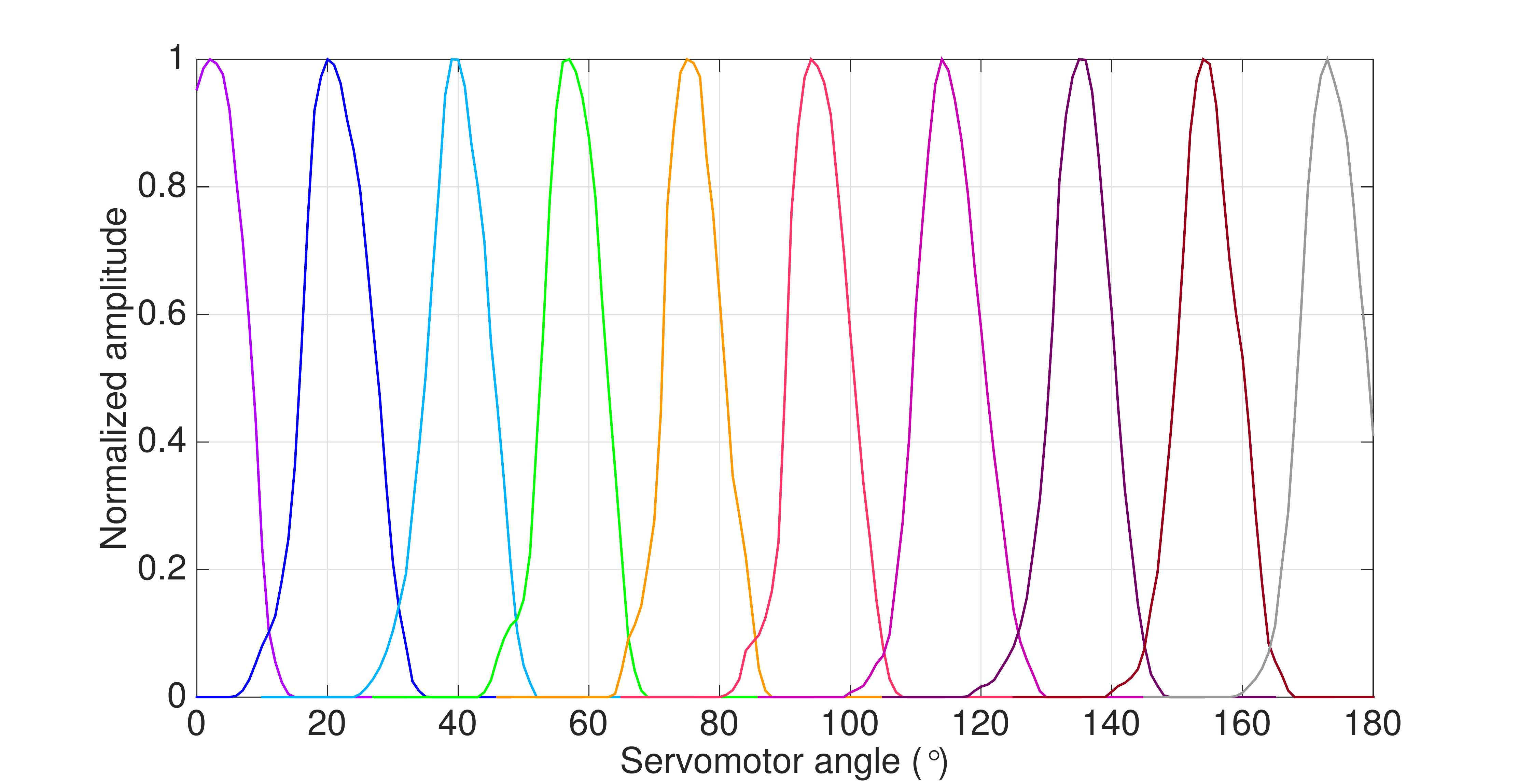}}
\caption{Normalized photodetector response as a function of LED servomotor angle position from $0-180$\degree. From right to left: 370\,nm, 470\,nm, 505\,nm, 591\,nm, 611\,nm, 624\,nm, 740\,nm, 770\,nm, 850\,nm, and 950\,nm LEDs. The LEDs were positioned on a LED board; rotating the LED board aligned the light from each LED with the collimating optics and onto the photodetector. We determined the servomotor angle that maximized the photodetector signal by sweeping the LED position in small increments around the estimated optimal angle.}
 \label{fig:MotorAngle}
\end{figure}

The full wiring diagram for the electronics in our device can be seen in Fig. \ref{fig:WiringDiag}. The photodetector response as a function of LED servomotor angle position from $0-180$\degree is shown in Fig. \ref{fig:MotorAngle}.

\begin{acknowledgements}

The authors would like to thank Hiruy Haile for help with machining components for the device, and thank Jose Romero and Sarah Kaiser for help with the electronics. We would also like to acknowledge John Curticapean for helpful advice concerning the rendering software, and to Matt Taylor and Jean-Philippe Bourgoin for their contributions with taking the spectrometer data of the LEDs, and to Jordan Coblin and Victor Reyes for their software contributions. The authors acknowledge funding from  NSERC, Industry Canada, Canadian Space Agency, CIFAR, CFI and ORF, and ONR.
\end{acknowledgements}

\end{document}